\documentclass[twocolumn]{revtex4-2}
\usepackage[english]{babel} 
\usepackage[utf8]{inputenc} 
\usepackage[a4paper, total={7in, 10in}]{geometry} 
\usepackage{graphicx} 
\usepackage{amsmath} 
\usepackage{empheq} 
\usepackage{bbold} 
\usepackage{physics} 
\usepackage[hidelinks]{hyperref} 
\usepackage{import} 
\usepackage[title]{appendix} 
\usepackage{xcolor} 
\usepackage{layouts}
\graphicspath{{.}}
\newtheorem{theorem}{Theorem}

\begin{document}

\newcommand{\PT}{$\mathcal{P}\mathcal{T}$}

\newcommand{\I}{\hat{\mathbb{1}}}
\newcommand{\sx}{\hat{\sigma}_x}
\newcommand{\sy}{\hat{\sigma}_y}
\newcommand{\sz}{\hat{\sigma}_z}

\newcommand{\Ifull}{\begin{bmatrix} 1 & 0 \\ 0 & 1 \end{bmatrix}}
\newcommand{\sxfull}{\begin{bmatrix} 0 & 1 \\ 1 & 0 \end{bmatrix}}
\newcommand{\syfull}{\begin{bmatrix} 0 & -i \\ i & 0 \end{bmatrix}}
\newcommand{\szfull}{\begin{bmatrix} 1 & 0 \\ 0 & -1 \end{bmatrix}}

\newcommand{\U}{\hat{U}}
\newcommand{\Hq}{\hat{H}_\mathrm{q}}
\newcommand{\Haq}{\hat{H}_\mathrm{aq}}
\newcommand{\Hop}{\hat{H}}
\newcommand{\Pop}{\hat{\mathcal{P}}}
\newcommand{\Top}{\hat{\mathcal{T}}}
\newcommand{\Cop}{\hat{\ast}}
\newcommand{\Mop}{\hat{M}}
\newcommand{\Gop}{\hat{\Gamma}}
\newcommand{\Lop}{\hat{\Lambda}}
\newcommand{\heta}{\hat{\eta}}
\newcommand{\p}[2]{P_{#1 \rightarrow #2}}
\newcommand{\Ugate}{\mathsf{U}(\theta, \phi, \lambda)}
\newcommand{\Ugateop}{\mathsf{\hat{U}}(\theta, \phi, \lambda)}
\newcommand{\iq}{_\mathrm{q}}
\newcommand{\ia}{_\mathrm{a}}
\newcommand{\iaq}{_\mathrm{aq}}

\newcommand{\vc}[1]{\boldsymbol{#1}}
\newcommand{\vop}[1]{\boldsymbol{\hat{#1}}}

\title{Quantum simulation of the pseudo-Hermitian Landau--Zener--Stückelberg--Majorana effect}

\author{Feliks Kivelä}
\author{Shruti Dogra}
\author{Gheorghe Sorin Paraoanu}
\affiliation{QTF  Centre  of  Excellence, Department of Applied Physics, Aalto University, FI-00076 Aalto, Finland}

\date{\today}


\begin{abstract}
While the Hamiltonians used in standard quantum mechanics are Hermitian, it is also possible to extend the theory to non-Hermitian Hamiltonians.
Particularly interesting are non-Hermitian Hamiltonians satisfying parity--time (\PT{}) symmetry, or more generally pseudo-Hermiticity, since such non-Hermitian Hamiltonians can still exhibit real eigenvalues.
In this work, we present a quantum simulation of the time-dependent non-Hermitian non-\PT{}-symmetric Hamiltonian used in a pseudo-Hermitian extension of the Landau--Zener--Stückelberg--Majorana (LZSM) model.
The simulation is implemented on a superconducting processor by using Naimark dilation to transform a non-Hermitian Hamiltonian for one qubit into a Hermitian Hamiltonian for a qubit and an ancilla; postselection on the ancilla state ensures that the qubit undergoes nonunitary time-evolution corresponding to the original non-Hermitian Hamiltonian.
We observe properties such as the dependence of transition rates on time and the replacement of conservation of total probability by other dynamical invariants in agreement with predictions based on a theoretical treatment of the pseudo-Hermitian LZSM system.
\end{abstract}

\maketitle


\section{Introduction}

Standard quantum mechanics is formulated in terms of Hermitian Hamiltonians, which guarantee real values for energies as well as unitary time evolution.
However, it is possible to extend the theory to non-Hermitian Hamiltonians in order to model open systems that significantly interact with their environment \cite{ashida}.
The study of open, effectively non-Hermitian quantum systems already began in the early days of quantum theory with works such as Gamow's use of a complex energy for an $\alpha$-decaying radioactive nucleus in 1928 \cite{gamow} and Majorana's theory of $\alpha$-particle absorption by a nucleus \cite{majorana}.
Subsequent early applications of non-Hermitian quantum theory included a derivation of the nuclear dispersion formula \cite{siegert} and a unified theory of nuclear reactions \cite{feshbach1958, feshbach1962}.

Non-Hermitian features are also a well-established part of fields such as superconductivity \cite{dynes78, dynes84} and X-ray absorption spectroscopy \cite{glatzel}, where an imaginary term $i\Gamma$ is often added to energies to model the phenomenon of lifetime broadening; i.e. the broadening of measured excitation peaks caused by the finite lifetime of the excited state.
In the context of superconductivity, $\Gamma$ is referred to as the Dynes parameter.

The above historical examples mainly feature non-Hermitian elements (such as the addition of imaginary terms to energies) as ad-hoc modifications of otherwise Hermitian standard quantum mechanics.
However, in 1998 Bender and Boettcher showed \cite{bender} that non-Hermitian Hamiltonians (NHHs) satisfying parity-inversion and time-reversal (\PT{}) symmetry can exhibit real eigenvalues (i.e. energies) despite their non-Hermiticity, which led to an increased interest in the study of explicitly non-Hermitian Hamiltonians.
In 2002, Mostafazadeh \cite{mostafazadeh} showed that \PT{} symmetry being associated with real eigenvalues is a consequence of \mbox{\PT{}-symmetric} Hamiltonians being a subset of the larger class of pseudo-Hermitian Hamiltonians, and that the pseudo-Hermiticity of a Hamiltonian is a necessary but not sufficient condition for it having real eigenvalues.

Interest into \PT{}-symmetric NHHs in particular has been driven by the fact that they can be experimentally realized as effective Hamiltonians in systems that involve a balanced gain-loss interaction with the environment \cite{ashida}.
These experimental realizations have been especially prominent in the field of optics \cite{el-ganainy, konotop}, where the Maxwell equations governing the dynamics of an electromagnetic wave in a waveguide under certain conditions can be written in a way that is identical with the Schrödinger equation describing a \mbox{\PT{}-s}ymmetric system, thus providing an analog to \mbox{\PT{}-s}ymmetric quantum mechanics \cite{ruschhaupt, guo, ruter, feng, regensburger}.
\mbox{\PT{}-s}ymmetric optics have been used to demonstrate exciting and exotic phenomena such as unidirectional invisibility \cite{lin, zhu, sounas, regensburger} and single-mode lasing \cite{gentry, hodaei2016, miri, hodaei, feng, hodaei2016b}.
Alongside optics, \PT{}-symmetric behaviour has also been realized on a large variety of other experimental platforms, such as the spins of ultracold $^6\text{Li}$ atoms \cite{li}, multilevel superconducting transmons \cite{naghiloo}, coupled LRC circuits \cite{schindler}, electromechanical resonators \cite{fleury}, trapped ions \cite{bian, cao}, $^{13}\text{C}$-labeled chloroform dissolved in \mbox{acetone-d6 \cite{lin_stochastic}}, optically-induced atomic lattices \cite{zhang}, and even macroscopic coupled \mbox{pendula \cite{bender_pendula}}. 

In addition to direct realizations, NHHs can also be studied by digital quantum simulation \cite{georgescu, paraoanu} by employing the Naimark dilation method \cite{kawabata}. This has been successfuly applied to platforms such as nitrogen-vacancy centres in diamond \cite{wu,zhang_topology,yu} and superconducting qubits \cite{dogra}.
However, all of these examples utilize a time-independent non-Hermitian Hamiltonian.

In this work we perform a quantum simulation of a time-dependent NHH. We extend the use of the Naimark dilation method to a time-dependent non-\PT{}-symmetric NHH in order to simulate the pseudo-Hermitian Landau--Zener--Stückelberg--Majorana (LZSM) model \cite{torosov} on a superconducting quantum processor. The LZSM model finds use in the description of transitions in a large variety of two-level systems, such as the valence/conduction bands in graphene \cite{higuchi}, molecular states in ultracold \mbox{$\text{Cs}_2$ \cite{mark}}, electron transfer between two As or P donors in a silicon nanowire transistor \cite{dupont-ferrier}, and even the in-plane and out-of-plane resonator modes of a classical nanomechanical resonator \cite{faust}.
The pseudo-Hermitian extension of the LZSM model can be applied to e.g. the description of two electromagnetic modes travelling to opposite directions in a waveguide \cite{yariv}, and boosting a weak signal with a stronger one in the sum-frequency generation process \cite{boyd}.

This article is organized as follows:
The next section (Section \ref{sec:lzsm}) presents a theoretical overview of the pseudo-Hermitian LZSM model.
Section \ref{sec:methods} then outlines the methods employed to simulate a non-Hermitian Hamiltonian on a quantum processor, and Section \ref{sec:results} presents the results of the application of this methodology to the simulation of the LZSM model described in Section \ref{sec:lzsm}.
Section \ref{sec:conclusion} ends the main article with a brief conclusion.
After the main article there are two Appendices: Appendix \ref{sec:appendix_pt_symmetry} presents the mathematical definitions of \PT{} symmetry and pseudo-Hermiticity, and \mbox{Appendix \ref{sec:appendix_lzsm}} includes calculations showing the application of these definitions to the pseudo-Hermitian LZSM model.


\section{The pseudo-Hermitian Landau--Zener--Stückelberg--Majorana model}
\label{sec:lzsm}

The pseudo-Hermitian Landau--Zener--Stückelberg--Majorana (LZSM) model \cite{torosov} features a Hamiltonian of the form
\begin{align}
    \label{eq:hamiltonian}
    \Hop(t)  &= \frac{1}{2} \begin{bmatrix} -\varepsilon(t) & \Omega_0 \\ k\Omega_0 & \varepsilon(t) \end{bmatrix} \\
    & = \frac{1}{2} \left[ \Omega_0 \left( \frac{k+1}{2}\sx - i \frac{k-1}{2}\sy \right) - \varepsilon(t) \sz \right],
\end{align}
where $\sigma_i$ are the Pauli matrices and $\varepsilon$ is varied linearly in time at a rate of $v$:
\begin{equation}
    \varepsilon(t) = vt.
\end{equation}
All parameters in the Hamiltonian are assumed to be real, with $v$ being positive.
The parameter $k$ controls the degree of non-Hermiticity; $k = 1$ reduces to the standard, Hermitian LZSM model \cite{ivakhnenko}.
For $k \neq 1$ the Hamiltonian is not Hermitian or \PT{}-symmetric, but is pseudo-Hermitian.
Mathematical definitions of \PT{} symmetry and pseudo-Hermiticity and details on how they apply to this Hamiltonian are included in Appendices \ref{sec:appendix_pt_symmetry} and \ref{sec:appendix_lzsm}.

At $\Omega_0 = 0$, the eigenstates of the system are the \textit{diabatic} states \cite{ivakhnenko}
\begin{equation}
    \ket{0} = \begin{bmatrix} 1 \\ 0 \end{bmatrix} \quad \text{and} \quad \ket{1} = \begin{bmatrix} 0 \\ 1 \end{bmatrix},
\end{equation}
with the energy levels
\begin{equation}
    E_0  = -\frac{\varepsilon}{2} \quad \text{and} \quad E_1  = \frac{\varepsilon}{2}.
\end{equation}
Adding a nonzero $\Omega_0$ gives rise to the \textit{adiabatic} energy states \cite{ivakhnenko}
\begin{align}
    \label{eq:adiabatic state}
    \ket{E_\pm} = \begin{bmatrix} -\varepsilon \pm \Delta E \\ k \Omega_0 \end{bmatrix}
\end{align}
with the corresponding eigenvalues
\begin{equation}
    \label{eq:adiabatic energy}
    E_\pm = \pm \frac{1}{2} \Delta E,
\end{equation}
where
\begin{equation}
    \Delta E = E_+ - E_- = \sqrt{k \Omega_0^2 + \varepsilon^2}
\end{equation}
is the difference between the energy levels.
Note that the form given in Eq. (\ref{eq:adiabatic state}) is not normalized.

\begin{figure}
    \centering
    \includegraphics[width=\columnwidth]{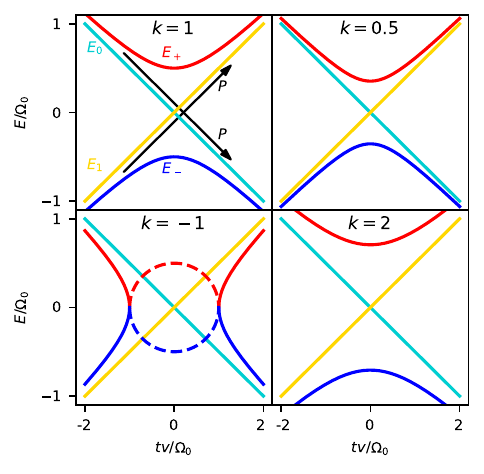} 
    \caption{A plot of the energies $E$ (in units of $\Omega_0$) of the diabatic and adiabatic states $\ket{0}$ (cyan), $\ket{1}$ (yellow), $\ket{E_+}$ (red), and $\ket{E_-}$ (blue) as a function of time (in units of $\Omega_0/v$), with various values of the parameter $k$.
    A system starting in the state $\ket{E_+}=\ket{0}$ (or $\ket{E_-}=\ket{1}$) at $t=-\infty$ and traversing the avoided crossing region towards $t=\infty$ will follow the diabatic state $\ket{0}$ ($\ket{1}$) and transition to $\ket{E_-}$ ($\ket{E_+}$) with a probability of $P$ (black arrows). Solid lines indicate purely real energy values, while the dashed lines in the $k=-1$ case represent the imaginary parts of purely imaginary energy values.}
    \label{fig:crossing}
\end{figure}

Fig. \ref{fig:crossing} presents the diabatic and adiabatic energy levels as a function of time.
For $t<0$, $\ket{1}$ is the ground state and $\ket{0}$ the excited state of the system, but at $t=0$ these levels cross, and for $t>0$, $\ket{1}$ is the excited state.
At large $|t|$ the diabatic and adiabatic levels approach each other, but at small $|t|$ the adiabatic energy levels diverge from the diabatic ones and exhibit an \textit{avoided crossing} near $t=0$ \cite{ivakhnenko}.

The question asked and answered by the LZSM model is the following: When time is scanned from $t=-\infty$ to $t=\infty$ across the avoided crossing region, what is the probability $P$ that these following, equivalent statements happen? \cite{ivakhnenko}
\begin{itemize}
    \item A system that starts in the diabatic state $\ket{0}$ or $\ket{1}$ at $t=-\infty$ will remain in the same state at $t=\infty$.
    \item A system that starts in the adiabatic state $\ket{E_+}$ or $\ket{E_-}$ at $t=-\infty$ will have transitioned to the other adiabatic state at $t=\infty$.
\end{itemize}
The answer is given by the LZSM formula \cite{ivakhnenko, torosov}
\begin{equation}
    P = P_{0 \rightarrow 0} = P_{1 \rightarrow 1} = \exp(-\pi\frac{k \Omega_0^2}{2 \hbar v}).
\end{equation}

As we see, increasing the value of the parameter $\Omega_0$ decreases $P$ (assuming $k > 0$ for simplicity).
$\Omega_0$ determines the smallest distance $\Delta E = \sqrt{k}|\Omega_0|$ (found at $t=0$) between the adiabatic energy levels, with a large $\Omega_0$ meaning the levels are far apart and transitions are unlikely.
For the diabatic levels, as the off-diagonal element of $\Hop$, $\Omega_0$ represents the coupling between the states $\ket{0}$ and $\ket{1}$, and a large $\Omega_0$ makes transitions between them more likely.

Counteracting the influence of $\Omega_0$ is the rate $v$ of the change in $\varepsilon(t)$.
If $\varepsilon$ changes slowly ($\hbar v/\Omega_0^2 \ll 1$), we have $P \approx 0$.
The system evolves \textit{adiabatically}; i.e. it mostly follows the adiabatic state $\ket{E_+}$ or $\ket{E_-}$.
With a fast rate of change ($\hbar v/\Omega_0^2 \gg 1$) we have $P \approx 1$, and the system undergoes \textit{diabatic} evolution where it likely transitions from $\ket{E_+}$ to $\ket{E_-}$ or vice versa by following the diabatic state $\ket{0}$ or $\ket{1}$ \cite{ivakhnenko}.

In the Hermitian case, the rates $P_{0 \rightarrow 1}$ and $P_{1 \rightarrow 0}$ of transitions between diabatic states are simply equal to $1 - P$, but in the pseudo-Hermitian case probability is not conserved and these instead take the forms \cite{torosov}
\begin{align}
    P_{0 \rightarrow 1} &= k(1 - P) \\
    P_{1 \rightarrow 0} &= \frac{1}{k}(1 - P).
\end{align}

The time-evolution of a system under the pseudo-Hermitian LZSM Hamiltonian presented in Eq. (\ref{eq:hamiltonian}) has been analytically solved \cite{torosov}, giving transition probabilities $\p{i}{f}(t)$ between states not just in the scenario where time evolves from $t_0=-\infty$ to $t=\infty$, but also for any finite length of time evolution.
The dynamical invariants of the system are given by the formulas
\begin{align}
    \label{eq:invariant0}
    k \p{0}{0}(t) + \p{0}{1}(t) &= k \\
    \label{eq:invariant1}
    k \p{1}{0}(t) + \p{1}{1}(t) &= 1,
\end{align}
which replace the conservation of total probability that is present in time-evolution under a Hermitian Hamiltonian \cite{simeonov, torosov}.

For the $k = -1$ case, the transition probabilities and the conservation law in Eq. (\ref{eq:invariant0}) have alse been presented by Malla et al. \cite{malla}, who investigated transition probabilities, invariants, and integrability in the context of anti-Hermitian LZSM systems.


\section{Methods}
\label{sec:methods}

\subsection{Time evolution operators}

The time evolution operator $\U(t, t_0)$ evolves an initial state $\ket{\psi(t_0)}$ to a final state $\ket{\psi(t)}$:
\begin{equation}
    \U(t, t_0) \ket{\psi(t_0)} = \ket{\psi(t)}.
\end{equation}
$\U(t, t_0)$ is determined by the Hamiltonian of the system as a \textit{time-ordered exponential}
\begin{align}
    \label{eq:U}
    \U(t, t_0) &= \hat{T} \exp(-\frac{i}{\hbar} \int_{t_0}^{t} \Hop(\tau) \dd \tau) \nonumber \\
    &\coloneqq 1 + \sum_{n=1}^{\infty} \frac{1}{n!} \left( -\frac{i}{\hbar} \right)^n \int_{t_0}^{t} \dd t_1 \int_{t_0}^{t} \dd t_2 \ldots \int_{t_0}^{t} \dd t_n \nonumber \\
    &\quad \quad \hat{T} \left( \Hop(t_1) \Hop(t_2) \ldots \Hop(t_n) \right),
\end{align}
where $\hat{T}$ is the \textit{time-ordering operator}
\begin{equation}
    \hat{T} \left( \hat{A}(t_1) \hat{B}(t_2) \right) = \begin{cases} \hat{A}(t_1) \hat{B}(t_2) & \text{if} \quad t_1 > t_2 \\ \hat{B}(t_2) \hat{A}(t_1) & \text{if} \quad t_2 > t_1, \end{cases}
\end{equation}
which rearranges a product between multiple operators (there can be more than two) in a decreasing order of time arguments when read from left to right.

If the Hamiltonian $\hat{H}(t)$ is Hermitian, $\U(t, t_0)$ is a unitary operator ($\U^\dagger\U=\I$), which means it preserves the inner product between states:
\begin{equation}
    \braket{\U \phi}{\U \psi} = \matrixelement{\phi}{\U^\dagger \U}{\psi} = \braket{\phi}{\psi}.
\end{equation}
A particular consequence of this is the conservation of probability: if the state $\ket{\psi}$ has been normalized so that
\begin{equation}
    \label{eq:normalized}
    \braket{\psi(t)}{\psi(t)} = 1
\end{equation}
at some time $t = t_0$, then Eq. (\ref{eq:normalized}) will also hold for all other values of $t$ as well.

\subsection{Simulating time evolution}

\subsubsection{Hermitian Hamiltonians}
A Hermitian Hamiltonian can be simulated on a quantum processor by computing (either numerically or analytically) the time evolution operator $\U(t_i, t_0)$ from \mbox{Eq. (\ref{eq:U})} for a range of time values $t_i$.
Each of these operators can then be implemented as a quantum circuit by decomposing it into a product of operators corresponding to quantum gates.
In practice we have used the \href{https://docs.quantum.ibm.com/api/qiskit/qiskit.circuit.QuantumCircuit}{\texttt{QuantumCircuit.unitary}} method from the Qiskit library \cite{qiskit} to perform this decomposition.

\begin{figure}
     \centering
     \includegraphics[trim={3.4cm 1cm 2.5cm 1cm}, clip, width=\columnwidth]{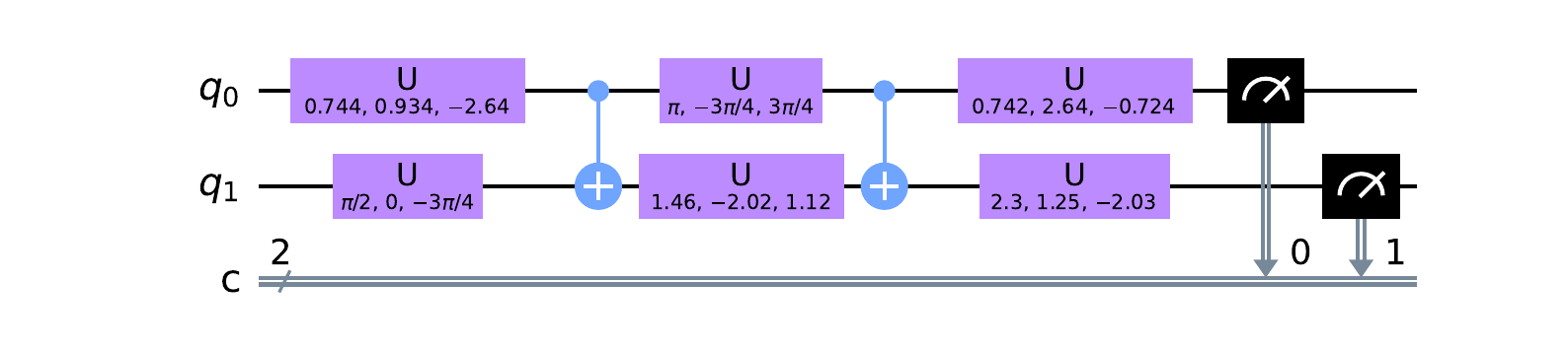}
     \caption{An example of a quantum circuit obtained by decomposing a time evolution operator (including state initialization as described in Eq.~(\ref{eq:initialization})). This particular circuit was used for the final data point (at $t = 20 \sqrt{\hbar/v}$) for the $k = 0.5$ case in the results presented below in Fig. \ref{fig:results0}. The circuit features two qubits ($q_0$ as the system qubit and $q_1$ as the ancilla) as well as a classical register (labeled ``c'') where the state of the system is measured to at the end of the circuit. The purple boxes are $\Ugate$ quantum gates that represent three-dimensional single-qubit rotations by the three given Euler angles. The blue symbols are controlled-NOT gates, with the smaller circle denoting the control qubit and the larger one the target qubit.}
     \label{fig:circuit}
\end{figure}

An example of a two-qubit quantum circuit is presented in Fig.~\ref{fig:circuit}.
The circuit consists of \href{https://docs.quantum.ibm.com/api/qiskit/qiskit.circuit.library.UGate}{$\Ugate$} quantum gates with \href{https://docs.quantum.ibm.com/api/qiskit/qiskit.circuit.library.CXGate}{controlled-NOT (CNOT)} gates between them.
A $\Ugate$ gate performs a three-dimensional one-qubit rotation by the Euler angles $\theta, \phi, \lambda$, and has a matrix representation of
\begin{equation}
    \Ugateop = \begin{bmatrix} \cos(\frac{\theta}{2}) & -e^{i\lambda} \sin(\frac{\theta}{2}) \\[5pt] e^{i\phi} \sin(\frac{\theta}{2}) & e^{i(\phi+\lambda)} \cos(\frac{\theta}{2}) \end{bmatrix}.
\end{equation}

When a quantum circuit that implements a time evolution operator is executed, its effect on the initial state $\ket{\psi(t_0)}$ of the qubits is equivalent to an operation by $\U(t_i, t_0)$, so that the state of the qubits at the end of the quantum circuit is $\U(t_i, t_0) \ket{\psi(t_0)} = \ket{\psi(t_i)}$.
Upon measurement each qubit is collapsed into one of the basis states $\ket{0}$ and $\ket{1}$.
Using a large number of shots (i.e. individual executions of a quantum circuit, each ending in a measurement) per circuit thus gives a good estimate of the probabilities of different measured states (which consist of $\ket{0}$ and $\ket{1}$ in a 1-qubit circuit, $\ket{00}$, $\ket{01}$, $\ket{10}$, and $\ket{11}$ in a 2-qubit one etc.).
When this process is repeated for all values $t_i$ in the selected range, data series describing the time evolution of the system in the form of probability functions $P_\mathrm{state}(t)$ are obtained.

\subsubsection{Non-Hermitian Hamiltonians}
For a non-Hermitian Hamiltonian, the simulation procedure described above cannot be used without alterations.
The time evolution operators $\U(t_i, t_0)$ produced by a non-Hermitian Hamiltonian can be nonunitary, and thus cannot be decomposed into a product of quantum gates, which only produce unitary transformations.
However, it is possible to simulate a Hermitian system which contains a subsystem that evolves in a non-Hermitian fashion.
More specifically, we use the mathematical technique of Naimark dilation to transform $\Hq$, a non-Hermitian Hamiltonian for one qubit, into $\Haq$, a Hermitian Hamiltonian for a qubit and an ancilla, so that the $\mathrm{ancilla} = 0$ subspace of the system evolves according to $\Hq$.

\subsection{Naimark dilation}

The Hermitian Hamiltonian $\Haq$, obtained from $\Hq$ as a result of Naimark dilation, can be expressed as \cite{wu, dogra}
\begin{equation}
    \Haq(t) = \I \otimes \Lop(t) + \sy \otimes \Gop(t), 
\end{equation}
where $\I$ is a $2 \times 2$ identity matrix and $\sy$ is the Pauli \mbox{$y$-matrix}.
The operators $\Lop(t)$ and $\Gop(t)$ are defined as
\begin{equation}
    \Lop(t) = \left[ \Hq(t) + i \dv{\heta(t)}{t} + \heta(t) \Hq(t) \heta(t) \right] \Mop^{-1}(t)
\end{equation}
and
\begin{equation}
    \Gop(t) = i \left[ \Hq(t) \heta(t) - \heta(t) \Hq(t) - i \dv{\heta(t)}{t} \right] \Mop^{-1}(t), 
\end{equation}
where
\begin{equation}
    \heta(t) = [\Mop(t) - \I]^\frac{1}{2},
\end{equation}
and
\begin{align}
    \Mop(t) = &\hat{T} \exp(-\frac{i}{\hbar} \int_{t_0}^{t} \Hq^\dagger(\tau) d \tau) \Mop(t_0) \nonumber \\ \times &\tilde{T} \exp(\frac{i}{\hbar} \int_{t_0}^{t} \Hq(\tau) d \tau).
\end{align}
In the last equation, $\tilde{T}$ is the \textit{time-anti-ordering operator} which arranges time arguments in an increasing instead of decreasing order.

The initial value $\Mop(t_0) = \Mop_0$ must be set so that $\Mop(t) - \I$ is positive for all $t$ in the simulated interval.
This is accomplished as follows: $\Mop_0$ is initially defined as
\begin{equation}
    \Mop'_0 = m_0 \times \I,
\end{equation}
where $m_0 > 1$ is an arbitrary constant.
Then the eigenvalues of $\Mop'(t)$ are obtained for all $t$ in the desired interval (in practice a discrete lattice of $t$-values is used).
The smallest eigenvalue (minimized over both the time axis and over the two different eigenvalues per time point) is labeled $\mu_\mathrm{min}$, and $\Mop'_0$ is replaced by the final value
\begin{equation}
    \Mop_0 = \frac{m_0}{\mu_\mathrm{min}} f \times \I,
\end{equation}
where $f > 1$ is another arbitrary constant.

Finally, in order to produce the desired time evolution for the qubit in the ancilla = 0 subspace, the qubit-ancilla system must be initialized in the correct initial state.
The qubit can be initialized in an arbitrary state; this state serves as the initial state for the non-Hermitian time-evolution.
However, the ancilla is initialized as $\ket{0}$ and then rotated around the $y$-axis by an angle $\theta$:
\begin{equation}
    \label{eq:ancilla_istate}
    \ket{\psi_0}\ia = \hat{R}_y(\theta) \ket{0}\ia, \quad \hat{R}_y(\theta) = e^{-i \theta \sy / 2}.
\end{equation}
The angle $\theta$ is given by
\begin{equation}
    \theta = 2 \atan \eta_0,
\end{equation}
with $\heta(t_0) = \eta_0 \times \I$; i.e. $\eta_0$ is the scalar equivalent of the matrix $\heta(t_0)$.

The rotation $\hat{R}_y(\theta)$ is combined with the time evolution operator $\U(t_i, t_0)$ before decomposition into a quantum circuit, so that the action of the quantum circuit upon the uninitialized state $\ket{00}_\mathrm{aq}$ can be mathematically expressed as
\begin{align}
    \label{eq:initialization}
    \ket{\psi(t_i)}\iaq &= \U(t_i, t_0)\iaq \nonumber \\
    &\times \left( \left[\hat{R}_y(\theta)\right]\ia \otimes \left[\hat{\mathrm{Init}}(\ket{\psi_0}\iq)\right]\iq \right) \ket{00}\iaq,
\end{align}
where $\hat{\mathrm{Init}}(\ket{\psi_0})$ is an operator that initializes the qubit to its given initial state $\ket{\psi_0}\iq$.
The lower indices a and q indicate whether a state or an operator applies to the ancilla, the qubit, or both. 


\section{Results}
\label{sec:results}

\subsection{Time evolution of probabilities}
\label{subsection:time_evolution}

\begin{figure}
     \centering
     \includegraphics[width=\columnwidth]{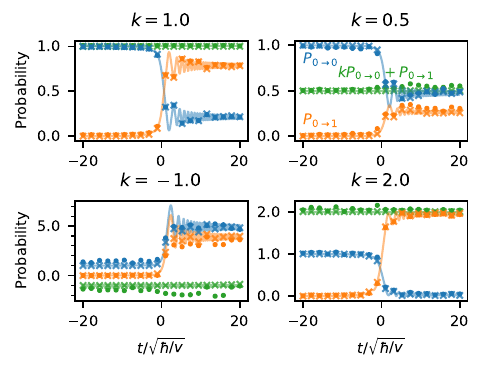}
     \caption{The probabilities of the states $\ket{0}\iq$ ($\p{0}{0}$, blue) and $\ket{1}\iq$ ($\p{0}{1}$, orange) as functions of time (in units of $\sqrt{\hbar/v}$), with $\ket{0}\iq$ as the initial state, $\Omega_0=\sqrt{\hbar v}$, and various values of $k$. The invariant $k \p{0}{0} + \p{0}{1}$ (green), which should be equal to $k$, is also included.
     Solid lines represent theoretical results, crosses are results obtained assuming a perfectly accurate quantum computer, and dots are results computed with a real quantum computer.}
     \label{fig:results0}
\end{figure}

The dilation procedure described above was applied to simulating the non-Hermitian LZSM model, with the aim of replicating theoretically predicted results \cite{torosov}.
Fig. 1 presents the results of the simulation of a qubit that starts in the state $\ket{0}\iq$ at $t_0 = -20 \sqrt{\hbar/v}$ and evolves until $t = 20 \sqrt{\hbar/v}$ under the Hamiltonian from \mbox{Eq. (\ref{eq:hamiltonian})}.
The blue line and markers indicate the probability $\p{0}{0}(t)$ of the qubit remaining in the initial state $\ket{0}\iq$ at a given time value, while the orange line and markers indicate the probability $\p{0}{1}(t)$ of the qubit having made the transition to $\ket{1}\iq$.

The solid lines represent a numerically computed theoretical prediction for the time evolution of the system, while the crosses represent the quantum circuits sent to be executed by the quantum processor.
Any difference between the solid lines and the crosses is an error caused by numerical inaccuracies in the Naimark dilation process and/or the decomposition of a time evolution operator into a quantum circuit.
The dots represent the results returned from the quantum processor; any difference between the crosses and the dots is an error introduced by the quantum processor, and is not present if a simulator emulating a perfectly accurate quantum processor is used in the place of a real quantum processor.

The simulation was performed on the \texttt{ibm\_lagos} processor, accessed through the \href{https://quantum.ibm.com/}{IBM Quantum Platform} cloud service.
Qubit $q_0$ of \texttt{ibm\_lagos} was used as the simulated non-Hermitian qubit, while qubit $q_1$ served as the ancilla.
Each data point was calculated with 10000 shots to ensure that the statistical uncertainties in the results are negligible. 

Running a quantum circuit on the processor returned values indicating how many times the system was found in each of the basis states $\ket{00}\iaq$, $\ket{01}\iaq$, $\ket{10}\iaq$, and $\ket{11}\iaq$ of the qubit-ancilla system.
This data corresponded to the time evolution of the Hermitian system with the Hamiltonian $\Haq$, and these basis states were equivalent to tensor products of the individual basis states of the qubit and the ancilla:
\begin{equation}
    \ket{ij}\iaq = \ket{i}\ia \otimes \ket{j}\iq, \quad i,j \in \{0,1\}.
\end{equation}

\begin{figure}
     \centering
     \includegraphics[width=\columnwidth]{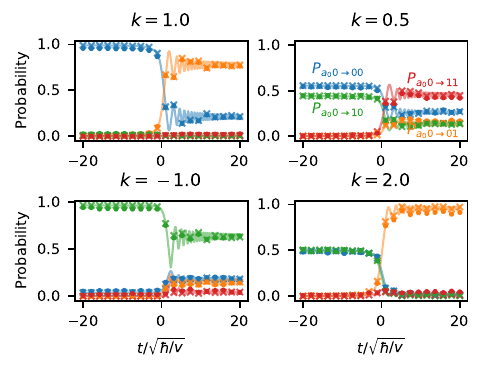}
     \caption{The data from Fig. \ref{fig:results0} before the postselection process: the probabilities of the states $\ket{00}\iaq$ ($\p{\mathrm{a}_00}{00}$, blue), $\ket{01}\iaq$ ($\p{\mathrm{a}_00}{01}$, orange), $\ket{10}\iaq$ ($\p{\mathrm{a}_00}{10}$, green), and $\ket{11}\iaq$ ($\p{\mathrm{a}_00}{11}$, red) as functions of time (in units of $\sqrt{\hbar/v}$), with $\Omega_0=\sqrt{\hbar v}$ and various values of $k$.
              The qubit starts in the state $\ket{0}\iq$, while the initial state of the ancilla (represented as a$_0$) is given by Eq. (\ref{eq:ancilla_istate}).}
     \label{fig:results4D}
\end{figure}

Data corresponding to the time evolution of the non-Hermitian system with the Hamiltonian $\Hq$ was acquired through postselecting for the $\mathrm{ancilla} = 0$ subspace.
The results for the $\mathrm{ancilla} = 0$ states $\ket{00}\iaq$ and $\ket{01}\iaq$ were used as those of the single-qubit subspace states $\ket{0}\iq$ and $\ket{1}\iq$ respectively, while the results for the \mbox{$\mathrm{ancilla} = 1$} states $\ket{10}\iaq$ and $\ket{11}\iaq$ were ignored.
Results from before the postselection process are presented in Fig. \ref{fig:results4D}.

Because the time evolution generated by $\Hq$ could be nonunitary, the probabilities $\p{0}{0}(t) = |\braket{0}{\psi(t)}\iq|^2$ and $\p{0}{1}(t) = |\braket{1}{\psi(t)}\iq|^2$ could be determined from the measured populations of the states $\ket{0}\iq$ and $\ket{1}\iq$ only up to a constant normalization factor $N_k$, which could be different for each value of $k$.
One possible way of determining $N_k$ was to note that in the theoretical model, $\p{0}{0}(t_0) + \p{0}{1}(t_0) = 1$ regardless of $k$, and set $1/ N_k$ equal to the sum of the measured populations of the states $\ket{0}\iq$ and $\ket{1}\iq$ at $t_0$. 
However, with this method any errors in the measured populations at $t_0$ would affect the measured probabilities $\p{0}{0}(t)$ and $\p{0}{1}(t)$ at all values of $t$.
This was particularly harmful in the $k=-1$ case, where the relative errors for the low-$t$ data points were large due to the very small size of the $\mathrm{ancilla} = 0$ subspace.
To solve this problem, optimal values for $N_k$ were determined through performing a least-squares fit of the measured results (dots in the figures) to the post-decomposition predicted results (crosses in the figures).

\subsection{Invariants}

\begin{figure}
    \centering
    \includegraphics[width=\columnwidth]{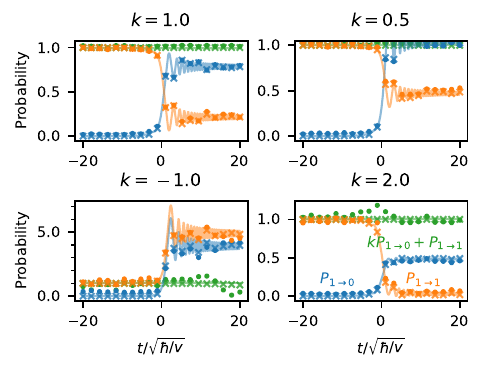}
    \caption{The probabilities of the states $\ket{0}\iq$ ($\p{1}{0}$, blue) and $\ket{1}\iq$ ($\p{1}{1}$, orange) as functions of time (in units of $\sqrt{\hbar/v}$), with $\ket{1}\iq$ as the initial state, $\Omega_0=\sqrt{\hbar v}$, and various values of $k$. The invariant $k \p{1}{0} + \p{1}{1}$ (green), which should be equal to 1, is also included.}
    \label{fig:results1}
\end{figure}

As can be seen in Fig. \ref{fig:results0}, in the non-Hermitian $k \neq 1$ cases the total probability $\p{0}{0} + \p{0}{1}$ changes as a function of time instead of being restricted to 1.
However, Eq. (\ref{eq:invariant0}) holds for all $k$ and $t$ (up to the limits of the accuracy of the quantum computer), and replaces the total probability as the invariant of the non-Hermitian system as predicted \cite{simeonov, torosov}. In order to show that the invariant from Eq. (\ref{eq:invariant1}) holds as well, Fig. \ref{fig:results0} was recreated with $\ket{1}\iq$ as the initial state; the results are presented in Fig. \ref{fig:results1}.

\begin{figure}
    \centering
    \includegraphics[width=\columnwidth]{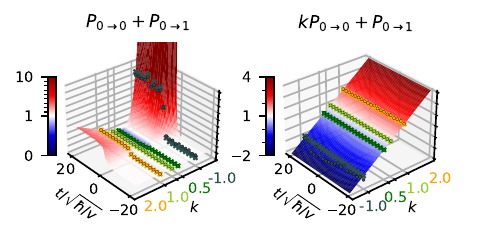}
    \caption{The dependence of the total probability $\p{0}{0} + \p{0}{1}$ and the invariant relation $k \p{0}{0} + \p{0}{1}$ on $k$ and time (in units of $\sqrt{\hbar/v}$), with $\Omega_0=\sqrt{\hbar v}$.}
    \label{fig:torosov_3d_plot}
\end{figure}

Fig. \ref{fig:torosov_3d_plot} presents the dependence of the total probability $\p{0}{0} + \p{0}{1}$ (left subplot) and the left side $k \p{0}{0} + \p{0}{1}$ of the invariant relation given in Eq. (\ref{eq:invariant0}) (right subplot) on the values of $k$ and time.
The three-dimensional surface represents numerically computed theoretical predictions, while the dots and crosses are based on the same data as the dots and crosses in Fig. \ref{fig:results0}.

\subsection{Dependence on $\Omega_0$}

\begin{figure}
    \centering
    \includegraphics[width=\columnwidth]{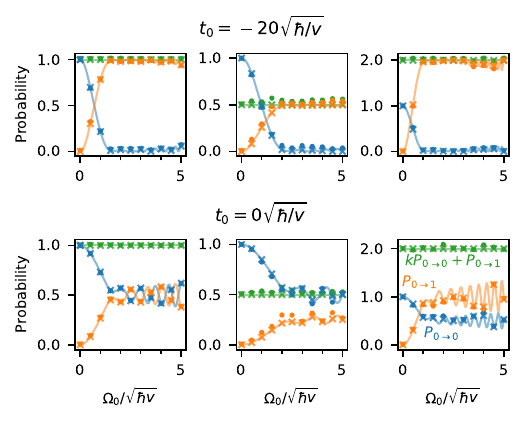}
    \caption{The probabilities of the states $\ket{0}\iq$ ($\p{0}{0}$, blue) and $\ket{1}\iq$ ($\p{0}{1}$, orange) as a function of $\Omega_0$ (in units of $\sqrt{\hbar v}$), with $\ket{0}\iq$ as the initial state, $t = 20 \sqrt{\hbar/v}$, and various values of $t_0$ and $k$. As $\Omega_0$ represents the coupling between the diabatic states, increasing its value increases the probability $\p{0}{1}$ of a transition.}
    \label{fig:rabi_results}
\end{figure}

Fig. \ref{fig:rabi_results} presents the dependence of the transition probabilities $\p{0}{0}$ and $\p{0}{1}$ on the parameter $\Omega_0$.
The $k = -1$ case has been omitted, because for that case the theoretical results feature probabilities as high as $10^{17}$, which were too large for the numerical algorithms used in the simulation.
The upper row features probabilities calculated with $t_0 = -20 \sqrt{\hbar/v}$ and $t = 20 \sqrt{\hbar/v}$, so each data point there corresponds to the final data point of a plot like those presented in Fig. \ref{fig:results0}, but with varying values of $\Omega_0$.
These data points have been measured and normalized in the manner explained in Section \ref{subsection:time_evolution}.
The lower row presents similar results but with $t_0 = 0 \sqrt{\hbar/v}$ and $t = 20 \sqrt{\hbar/v}$ instead.


\section{Conclusion}
\label{sec:conclusion}

We have realized a quantum simulation of a single qubit under a time-dependent pseudo-Hermitian LZSM Hamiltonian, and used the resulting data to validate previous theoretical predictions.
We have observed transition rates and the nonconservation of probability, the dependence of the results on time and the parameters $k$ and $\Omega_0$, and the presence of the dynamical invariants of the system.
The quantum simulation employs the Naimark dilation technique; by extending the Hilbert space with the use of an ancilla qubit and postselecting on the ancilla state, we were able to implement specific non-Hermitian dynamics.

This work demonstrates a useful application of current quantum computing technology, and shows that simulation of time-dependent non-Hermitian Hamiltonians is possible on a superconducting quantum processor.
Future research will be focused on extending the methodology presented in this article to systems of more than one qubit, which will yield insights into multi-qubit phenomena such as quantum entanglement in a non-Hermitian context.

\acknowledgments

FK acknowledges funding from the Magnus Ehrnrooth foundation and the Nokia Industrial Doctoral School in Quantum Technology.
SD and GSP acknowledge support from the Academy of Finland (Research Council of Finland) through the Centre of Excellence program (project 352925).

\clearpage
\onecolumngrid
\begin{appendices}


\section{\PT{} symmetry and pseudo-Hermiticity}
\label{sec:appendix_pt_symmetry}

\subsection{\PT{} symmetry}

A Hamiltonian is considered \PT{}-symmetric if it commutes with the $\Pop \Top$ operator ($[\Hop, \Pop \Top] = 0$), which is a combination of the parity-inversion operator $\Pop$ and the time-reversal operator $\Top$.
The parity-inversion operator $\Pop$ produces a mirror image of a system by inverting the signs of all positions $\vop{x}$ and momenta $\vop{p}$ on its right \mbox{side \cite{ashida, bender_basics, bender_intro}}:
\begin{equation}
    \Pop \vop{x} = -\vop{x} \Pop, \quad \Pop \vop{p} = -\vop{p} \Pop.
\end{equation}
The time-reversal operator $\Top$ reverses the flow of time by inverting the signs of momenta and replacing every $i$ with a $-i$ (i.e. performing complex conjugation) \cite{ashida, bender_basics, bender_intro}:
\begin{equation}
    \Top \vop{x} = \vop{x} \Top, \quad \Top \vop{p} = -\vop{p} \Top, \quad \Top i = -i \Top.
\end{equation}

In the simplest case of a two-state system, $\Pop$ is equal to the Pauli matrix $\sx$, and $\Top$ performs complex conjugation to everything on its right side \cite{bender_intro, bender_pendula, dogra}.
The generic form for a \PT{}-symmetric two-state Hamiltonian is \cite{bender_generalized}
\begin{equation}
    \label{eq:generic}
    \Hop = \begin{bmatrix} a & b \\ b^* & a^* \end{bmatrix},
\end{equation}
where $a$ and $b$ are complex parameters.
The Hamiltonian can also be expressed in terms of Pauli matrices as 
\begin{align}
    \label{eq:generic pauli}
    r_1 \I + r_x \sx + r_y \sy + i r_z \sz,
\end{align}
where $r_i$ are real parameters.

A remarkable property of \PT{}-symmetric Hamiltonians is that of the symmetry existing in either a \textit{broken} or an \textit{unbroken} state.
\PT{} symmetry is considered unbroken when every eigenstate of the Hamiltonian is also an eigenstate of the $\Pop \Top$ operator.
If this is the case, the spectrum (i.e. the multiset of the eigenvalues) of the Hamiltonian is entirely real.
The symmetry is considered broken if some of the eigenstates of the Hamiltonian are not eigenstates of the $\Pop \Top$ operator; in this case some eigenvalues appear as pairs of values that are complex conjugate to each \mbox{other \cite{ashida, bender_intro}}.

If the Hamiltonian depends on a parameter, changing the value of this parameter can move the Hamiltonian from a state of broken \PT{} symmetry to a state of unbroken symmetry, or vice versa.
This \textit{\PT{} phase transition} typically occurs at an \textit{exceptional point} in the parameter space, where the different eigenvectors become parallel and the eigenvalues are equal \cite{ashida, el-ganainy}.

\subsection{Pseudo-Hermiticity}

While \PT{}-symmetric Hamiltonians are often associated with real eigenspectra, \PT{} symmetry is in fact neither a sufficient nor a necessary condition for a real spectrum.
In general, real eigenspectra are connected to the concept of pseudo-Hermitian Hamiltonians, of which \PT{}-symmetric Hamiltonians constitute a particularly simple subclass.
An operator $\Hop$ is considered \textit{$\heta$-pseudo-Hermitian} if 
\begin{equation}
    \label{eq:pseudohermiticity}
    \Hop^\dagger = \heta \Hop \heta^{-1},
\end{equation}
where $\heta$ is an invertible Hermitian operator. 
In general, $\Hop$ is considered \textit{pseudo-Hermitian} if an operator $\heta$ exists so that Eq. (\ref{eq:pseudohermiticity}) holds.
Generally $\heta$ is not unique; there can be several different $\heta$ which can satisfy Eq. (\ref{eq:pseudohermiticity}) for a given Hamiltonian.
Note that Hermiticity is a special case of pseudo-Hermiticity, since a Hermitian operator is also $\I$-pseudo-Hermitian \cite{mostafazadeh, ashida}.

All Hamiltonians with real eigenvalues are pseudo-Hermitian, but a given pseudo-Hermitian Hamiltonian does not necessarily have a real spectrum.
The following two theorems apply to a Hamiltonian $\Hop$ with a complete biorthonormal eigenbasis and a discrete spectrum and describe the conditions under which a real spectrum is present:
\begin{theorem}
\label{thrm:pseudo-hermitian}
\textup{\cite{mostafazadeh, ashida}}
$\Hop$ is pseudo-Hermitian if and only if one of the following conditions holds:
\begin{itemize}
    \item The spectrum of $\Hop$ is real.
    \item The eigenvalues of $\Hop$ appear in complex conjugate pairs and the multiplicities of complex conjugate eigenvalues (i.e. how many degenerate eigenvalues correspond to an eigenvector) are the same.
\end{itemize}
\end{theorem}
\begin{theorem}
\label{thrm:spectrum}
\textup{\cite{mostafazadeh2, ashida}}
$\Hop$ has a real spectrum if and only if there is an invertible linear operator $\hat{O}$ such that $\Hop$ is $\hat{O}\hat{O}^\dagger$-pseudo-Hermitian; i.e. $\Hop^\dagger = \hat{O}\hat{O}^\dagger \Hop \left(\hat{O}\hat{O}^\dagger\right)^{-1}$.
\end{theorem}
The requirement of a complete biorthonormal eigenbasis is equivalent to stating that the Hamiltonian is diagonizable.
This is not a very strict requirement, as practically all physically relevant Hamiltonians fulfill it \cite{mostafazadeh3}.


\section{Mathematical details on the pseudo-Hermitian LZSM Hamiltonian}
\label{sec:appendix_lzsm}

\subsection{\PT{} symmetry}
\subsubsection{Original Hamiltonian}

The pseudo-Hermitian LZSM Hamiltonian from Eq. (\ref{eq:hamiltonian}) is not \PT{}-symmetric, as it does not match the generic forms given in Eqs. (\ref{eq:generic}) and (\ref{eq:generic pauli}).
An explicit calculation of the commutator $[\Hop, \Pop \Top]$ yields
\begin{align}
    &[\Hop, \Pop \Top] = \Hop \Pop \Top - \Pop \Top \Hop \nonumber \\
    &= \frac{1}{2} \begin{bmatrix} -\varepsilon & \Omega_0 \\ k \Omega_0 & \varepsilon \end{bmatrix} \begin{bmatrix} 0 & 1 \\ 1 & 0 \end{bmatrix} \hat{\ast} - \begin{bmatrix} 0 & 1 \\ 1 & 0 \end{bmatrix} \hat{\ast} \frac{1}{2} \begin{bmatrix} -\varepsilon & \Omega_0 \\ k \Omega_0 & \varepsilon \end{bmatrix} \nonumber \\
    &= \frac{1}{2} \begin{bmatrix} (1-k) \Omega_0 & -2\varepsilon \\ 2\varepsilon & (k-1) \Omega_0 \end{bmatrix} \hat{\ast},
\end{align}
where $\hat{\ast}$ is an operator that performs complex conjugation on everything to its right side.
The result further confirms the lack of \PT{} symmetry of the Hamiltonian.
In the special case of $k = 1, \: t = 0$, the Hamiltonian is reduced to $\Omega_0 \sx / 2$, which is \PT{}-symmetric.

\subsubsection{Rotated Hamiltonian}

While the Hamiltonian from Eq. (\ref{eq:hamiltonian}) is not \PT{}-symmetric, it can be transformed into a \PT{}-symmetric one by a $\pi / 2$ rotation around the $x$-axis.
This rotation transforms the Pauli matrices as $\sx \rightarrow \sx$, $\sy \rightarrow \sz$, and $\sz \rightarrow - \sy$, so that Eq. (\ref{eq:hamiltonian}) is transformed into
\begin{align}
    \label{eq:rotated hamiltonian}
    \Hop_\mathrm{rot} &= \frac{1}{2} \left[  \frac{k+1}{2} \Omega_0 \sx  + \varepsilon \sy - i \frac{k-1}{2} \Omega_0 \sz \right] = \frac{1}{2} \begin{bmatrix} -i \frac{k-1}{2} \Omega_0 & \frac{k+1}{2} \Omega_0 -i \varepsilon \\ \frac{k+1}{2} \Omega_0 +i \varepsilon & i \frac{k-1}{2} \Omega_0 \end{bmatrix},
\end{align}
which matches the forms given in Eq. (\ref{eq:generic}) and Eq. (\ref{eq:generic pauli}).
The commutator is
\begin{align}
    [\Hop_\mathrm{rot}, \Pop \Top]
    =& \frac{1}{2} \left( \begin{bmatrix} -i \frac{k-1}{2} \Omega_0 & \frac{k+1}{2} \Omega_0 -i \varepsilon \\ \frac{k+1}{2} \Omega_0 +i \varepsilon & i \frac{k-1}{2} \Omega_0 \end{bmatrix} \sxfull - \sxfull \begin{bmatrix} i \frac{k-1}{2} \Omega_0 & \frac{k+1}{2} \Omega_0 + i \varepsilon \\ \frac{k+1}{2} \Omega_0 - i \varepsilon & -i \frac{k-1}{2} \Omega_0 \end{bmatrix} \right) \Cop = 0,
\end{align}
which explicitly shows the \PT{} symmetry.

\subsection{Eigenvalues and eigenvectors}
\subsubsection{Original Hamiltonian}

The following calculation shows that the expressions given in Eqs. (\ref{eq:adiabatic state}) and (\ref{eq:adiabatic energy}) are indeed the eigenvectors and eigenvalues of the Hamiltonian $\Hop$ from Eq. (\ref{eq:hamiltonian}):
\begin{align}
    \Hop \ket{E_\pm} &= \frac{1}{2} \begin{bmatrix} -\varepsilon & \Omega_0 \\ k \Omega_0 & \varepsilon \end{bmatrix} \begin{bmatrix} -\varepsilon \pm \Delta E \\ k \Omega_0 \end{bmatrix}
    = \frac{1}{2} \begin{bmatrix} \mp \varepsilon \Delta E + \smash{\overbrace{\varepsilon^2 + k \Omega_0^2}^{=(\Delta E)^2}} \\ \pm k \Omega_0 \Delta E \end{bmatrix}
    = \underbrace{\pm \frac{1}{2} \Delta E}_{=E_\pm} \underbrace{\begin{bmatrix} -\varepsilon \pm \Delta E \\ k \Omega_0 \end{bmatrix}}_{=\ket{E_\pm}}.
\end{align}

\subsubsection{Rotated Hamiltonian}

For the rotated Hamiltonian $\Hop_\mathrm{rot}$ from Eq. (\ref{eq:rotated hamiltonian}), the eigenvalues are the same as given in Eq. (\ref{eq:adiabatic energy}), but the eigenvectors change to
\begin{equation}
    \ket{E_\pm}_\mathrm{rot} = \begin{bmatrix} \pm \Delta E - i \frac{k-1}{2} \Omega_0 \\ \frac{k+1}{2} \Omega_0 + i \varepsilon \end{bmatrix},
\end{equation}
since
\begin{align}
    \Hop_\mathrm{rot} \ket{E_\pm}_\mathrm{rot} &= \frac{1}{2} \begin{bmatrix} -i \frac{k-1}{2} \Omega_0 & \frac{k+1}{2} \Omega_0 -i \varepsilon \\ \frac{k+1}{2} \Omega_0 +i \varepsilon & i \frac{k-1}{2} \Omega_0 \end{bmatrix} \begin{bmatrix} \pm \Delta E - i \frac{k-1}{2} \Omega_0 \\ \frac{k+1}{2} \Omega_0 + i \varepsilon \end{bmatrix} \nonumber
    = \frac{1}{2} \begin{bmatrix} \mp i\frac{k-1}{2}\Omega_0\Delta E + \smash{\overbrace{k\Omega_0^2 + \varepsilon^2}^{=(\Delta E)^2}} \\[2pt] \pm\frac{k+1}{2}\Omega_0\Delta E \pm i\varepsilon\Delta E \end{bmatrix} \nonumber \\
    &= \underbrace{\pm \frac{1}{2} \Delta E}_{=E_\pm} \underbrace{\begin{bmatrix} \pm \Delta E - i\frac{k-1}{2}\Omega_0 \\ \frac{k+1}{2}\Omega_0 + i\varepsilon \end{bmatrix}}_{=\ket{E_\pm}_\mathrm{rot}}.
\end{align}

The eigenvalues are real when $k\Omega_0^2+\varepsilon^2 \geq 0$, and appear as imaginary complex conjugate pairs when $k\Omega_0^2+\varepsilon^2 < 0$.
$k\Omega_0^2+\varepsilon^2 = 0$ is an exceptional point where the eigenvalues and eigenvectors of both $\Hop$ and $\Hop_\mathrm{rot}$ become degenerate.
Based on the realness of eigenvalues, $k\Omega_0^2+\varepsilon^2 > 0$ can be assumed to be the region of unbroken \PT{} symmetry for $\Hop_\mathrm{rot}$, with $k\Omega_0^2+\varepsilon^2 < 0$ corresponding to broken \PT{} symmetry.
This can be confirmed by computing the effect of the \PT{} operator on the eigenstates $\ket{E_\pm}_\mathrm{rot}$ of $\Hop_\mathrm{rot}$:
\begin{align}
    \label{eq:PT on Psi_rot}
    \Pop\Top \ket{E_\pm}_\mathrm{rot}
    &= \sxfull \begin{bmatrix} \pm \Delta E - i\frac{k-1}{2}\Omega_0 \\ \frac{k+1}{2}\Omega_0 + i\varepsilon \end{bmatrix}^*
    = \begin{bmatrix} \dfrac{\frac{k+1}{2}\Omega_0 - i\varepsilon}{\pm \Delta E -i\frac{k-1}{2}\Omega_0} \left( \pm \Delta E - i\frac{k-1}{2}\Omega_0 \right) \\[15pt] \dfrac{\pm (\Delta E)^* + i\frac{k-1}{2}\Omega_0}{\frac{k+1}{2}\Omega_0 + i \varepsilon} \left( \frac{k+1}{2}\Omega_0 + i \varepsilon \right) \end{bmatrix}.
\end{align}
Here we can transform the fraction in the first element of the vector into
\begin{align}
    \dfrac{\frac{k+1}{2}\Omega_0 - i\varepsilon}{\pm \Delta E -i\frac{k-1}{2}\Omega_0}
    &= \dfrac{\left(\frac{k+1}{2}\Omega_0 - i\varepsilon\right)\left(\frac{k+1}{2}\Omega_0 + i\varepsilon\right)}{\left(\pm \Delta E -i\frac{k-1}{2}\Omega_0\right)\left(\frac{k+1}{2}\Omega_0 + i\varepsilon\right)}
    = \frac{\pm\Delta E + i\frac{k-1}{2}\Omega_0}{\frac{k+1}{2}\Omega_0 + i\varepsilon},
\end{align}
since
\begin{align}
    \left(\frac{k+1}{2}\Omega_0 - i\varepsilon\right)\left(\frac{k+1}{2}\Omega_0 + i\varepsilon\right)
    = \left(\frac{k-1}{2}\Omega_0\right)^2 + \underbrace{k\Omega_0^2 + \varepsilon^2}_{=(\Delta E)^2}
    = \left(\pm\Delta E + i\frac{k-1}{2}\Omega_0\right)\left(\pm\Delta E - i\frac{k-1}{2}\Omega_0\right).
\end{align}
In the region where $\Delta E \in \mathbb{R} \Leftrightarrow k\Omega_0^2+\varepsilon^2 \geq 0$, the complex conjugation in the second element of the vector in \mbox{Eq. (\ref{eq:PT on Psi_rot})} has no effect, and we have
\begin{align}
    \label{eq:PT eigenvalue}
    \Pop\Top \ket{E_\pm}_\mathrm{rot} &= \dfrac{\pm \Delta E + i\frac{k-1}{2}\Omega_0}{\frac{k+1}{2}\Omega_0 + i \varepsilon} \underbrace{\begin{bmatrix} \pm \Delta E - i\frac{k-1}{2}\Omega_0 \\ \frac{k+1}{2}\Omega_0 + i \varepsilon \end{bmatrix}}_{=\ket{E_\pm}_\mathrm{rot}}.
\end{align}
This shows that the eigenstates of $\Hop_\mathrm{rot}$ are also eigenstates of $\Pop\Top$, which is the condition for unbroken \PT{} symmetry.

\subsection{Pseudo-Hermiticity}

The Hamiltonian $\Hop$ from Eq. (\ref{eq:hamiltonian}) can be easily seen to be $\heta$-pseudo-Hermitian with
\begin{equation}
    \label{eq:eta}
    \heta = a \begin{bmatrix} k & 0 \\ 0 & 1 \end{bmatrix} = a\frac{k+1}{2} \I + a\frac{k-1}{2} \sz,
\end{equation}
where $a$ is a real constant, since
\begin{align}
    \heta \Hop \heta^{-1} &= a \begin{bmatrix} k & 0 \\ 0 & 1 \end{bmatrix} \frac{1}{2} \begin{bmatrix} -\varepsilon & \Omega_0 \\ k \Omega_0 & \varepsilon \end{bmatrix} \frac{1}{a} \begin{bmatrix} \frac{1}{k} & 0 \\ 0 & 1 \end{bmatrix}
    = \frac{1}{2} \begin{bmatrix} -\varepsilon & k \Omega_0 \\ \Omega_0 & \varepsilon \end{bmatrix} = \Hop^\dagger.
\end{align}
Note that this choice of $\heta$ is a particularly simple solution, but not a unique one.

\end{appendices}
\twocolumngrid
\bibliography{./bibliography.bib}

\end{document}